\documentstyle[eqsecnum,aps,prd,epsf]{revtex}
\def\sqr#1#2{{\vcenter{\hrule height.#2pt
      \hbox{\vrule width.#2pt height#1pt \kern#1pt
          \vrule width.#2pt}
      \hrule height.#2pt}}}

\def\edth{{\rlap{$\partial$}\raise0.3em\hbox{$-$}}}

\def\gtrless{{\hbox{\raisebox{0.6ex}{$\,>$}}\kern-1.8ex
                   \hbox{\raisebox{-0.6ex}{$<\,$}}}}

\begin{document}
\hspace{-10mm}
\leftline{\epsfbox{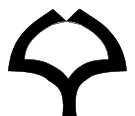}}
\vspace{-10.0mm} 
\thispagestyle{empty}
{\baselineskip-4pt
\font\yitp=cmmib10 scaled\magstep2
\font\elevenmib=cmmib10 scaled\magstep1  \skewchar\elevenmib='177
\leftline{\baselineskip20pt
\hspace{10mm} 
\vbox to0pt
   { {\yitp\hbox{Osaka \hspace{1.5mm} University} }
     {\large\sl\hbox{{Theoretical Astrophysics}} }\vss}
}
\rightline{
\large\baselineskip20pt\rm\vbox to20pt{
\baselineskip14pt
\hbox{OU-TAP-156}
\hbox{UTAP-385}
\vspace{1mm}
\hbox{\today}\vss}}%
}
\vspace{8mm}

\begin{center}

{\Large \bf Self-Force on a Scalar Charge
in Circular Orbit\\ around a Schwarzschild Black Hole}

\bigskip

Hiroyuki Nakano$^{1}$, Yasushi Mino$^{2}$ and Misao Sasaki$^{1}$

\smallskip

$^1${\em Department of Earth and Space Science,~Graduate School of
  Science,~Osaka University,\\ Toyonaka, Osaka 560-0043, Japan
}\\
\smallskip
$^2${\em Theoretical Astrophysics, California Institute of Technology,
Pasadena,~California,~91125, USA}\\

\bigskip

\end{center}

\begin{center}

{\bf abstract}

\end{center}

{\small
In an accompanying paper,
we have formulated two types of regularization methods
to calculate the scalar self-force
on a particle of charge $q$ moving around a black hole of
 mass $M$ \cite{MNS1}, one of which is called
the ``power expansion regularization''.
In this paper, we analytically evaluate the self-force
(which we also call the reaction force) to
the third post-Newtonian (3PN) order
on the scalar particle in circular orbit around
a Schwarzschild black hole by using the power expansion regularization.
It is found that the $r$-component of the self-force arises
at the 3PN order, whereas the $t$- and $\phi$-components,
which are due to the radiation reaction, appear
at the 2PN and 1.5PN orders, respectively.
}

\section{Introduction}

Thanks to the recent technological advance,
we have almost come to the stage that gravitational waves
are detectable. The observation of gravitational waves is
absolutely a new window to observe our universe.
We also expect that the observation of gravitational waves
provides a direct experimental test of general relativity.

There are several on-going projects of gravitational wave detection
in the world, such as LIGO\cite{LIGO}, VIRGO\cite{VIRGO},
GEO\cite{GEO} and TAMA\cite{TAMA}.
In addition, LISA\cite{LISA} is proposed a future project.
LISA is unique in that it is a space-based observation.
Since the detector will be launched in space,
it will be free from the seismic noise,
and it will be remarkably sensitive
to the low frequency gravitational waves below 1 Hz.
In contrast to the ground-based gravitational wave detectors
whose main targets are inspiralling binaries of
solar-mass compact objects (either black holes or neutron stars),
LISA will detect the gravitational waves from a solar-mass compact
object orbiting a supermassive black hole
at galactic centers. In the latter situation, the
radiation reaction effect over a few orbital periods will
be much smaller than the case of compact binaries because of
the large mass ratio between the black hole and the compact object.
However, the accumulated ``secular" effect of radiation reaction
will be as important as in the case of compact binaries.
Furthermore, the spacetime around a supermassive black hole
is expected to be described by the Kerr geometry
 with non-negligible spin. Hence
the relativistic effect, especially the effect of the
non-spherical geometry, will significantly affect
the orbital evolution.
In this situation, the calculation of the orbital evolution
by means of the standard post-Newtonian expansion method
seems formidable to perform.
Instead, the black hole perturbation approach seems much more relevant,
where the compact object is approximated by a particle orbiting
a Kerr black hole (and perturbing the geometry).

For a binary system of almost equal mass objects, the orbit is
expected to become sufficiently circular at the final inspiral
stage ($\sim 10^5$ cycles before coalescence)
by the radiation reaction. However, this will not be the case
for a binary systems of extreme mass ratio, and the orbit may
not become circular until the very last moment.
Thus we are required to consider general orbits.

Previous work on radiation reaction in the black hole perturbation
approach has been based on conservation of the total energy and angular
momentum of the system (see Ref.~\cite{MSSTT} for a review).
However, this approach cannot work
for general orbits around a Kerr black hole,
since the orbit is characterized by the three parameters;
the energy, angular momentum (with respect to the symmetry axis),
and the Carter constant \cite{Carter}. Since the Carter constant
is not associated with the Killing vector field of the
Kerr geometry, its evolution cannot be determined by the conservation
laws. Instead, it is necessary to derive the reaction force
acting locally on the orbit.

Recently, the equation of motion of a point particle on a
curved background with the effect of gravitational radiation reaction
was derived by Mino, Sasaki and Tanaka
by solving the Einstein equations by means of
matched asymptotic expansion \cite{reaction},
and by Quinn and Wald by an axiomatic approach \cite{QuiWal}.

The essential problem of the equation of motion of a point particle
was that the metric perturbation induced by the particle diverges
at the location of the particle.
Apparently, divergence of the metric perturbation means the
break down of the linear approximation. Nevertheless,
we may expect that the strong equivalence principle still holds
to an extent, and that the notion of the equation of motion of
a point particle is still useful.
The work of Ref.~\cite{reaction} and Ref.~\cite{QuiWal}
has shown this is indeed the case, and
that the extra force by particle's self-gravity can be derived
from the linear metric perturbation by subtracting the singular part
appropriately.

The singular part irrelevant to the motion can be exactly
identified by the covariant Hadamard method \cite{Had}
as the coincidence limit of the part that directly
propagates to field points along the background null cone of the source.
The rest of the metric perturbation is called the tail part,
which propagates within the null cone due to the curvature
scattering.
In practice, the direct part can be calculated by using
the method of DeWitt and Brehme \cite{DeWBre}.
On the other hand, the calculation of the tail part
requires full knowledge of the metric perturbation,
and it is impossible to calculate it for a general spacetime.
Fortunately, in the case of the black hole perturbation,
it is in principle possible to calculate
the full metric perturbation \cite{Chrz} by using the Regge-Wheeler-Teukolsky
formalism \cite{Teuk}. Then we may extract the tail part by
subtracting the direct part from the full metric perturbation.

Here, however, we must mention a couple of non-trivial
obstacles when we attempt to perform the extraction procedure:
\begin{list}{}{}
  \item[1)] Subtraction problem\\
In the Regge-Wheeler-Teukolsky formalism,
the full metric perturbation is calculated in the form of
Fourier-harmonic series.\footnote
{Throughout the paper, the Fourier modes refer to
$e^{-i\omega t}$ and the harmonic modes to $Y_{\ell m}(\theta,\phi)$.}
 This fact totally obscures the local spacetime
behavior of the metric perturbation around the particle.
 On the other hand, the direct part is given
only locally around the location of the particle. This nature
of locality makes it very difficult for us to transform the
direct part to the Fourier-harmonic series form.
Thus the subtraction of the direct part from the full metric perturbation
is a highly non-trivial task.
\item[2)] Gauge problem\\
The Hadamard prescription which identifies the direct part
is, by definition, applicable only to hyperbolic differential equations,
and specifically to the perturbation equations in the
Lorentz (harmonic) gauge in the case of our interest.
On the other hand, the metric perturbation directly obtainable in
the Regge-Wheeler-Teukolsky formalism is not the one in the
Lorentz gauge. Hence an appropriate gauge transformation
should be worked out.
\end{list}

In addition to the above, in the case of a Kerr background,
which is of our ultimate interest, there is a much more intricate
technical issue of how to treat the spheroidal harmonics:
Neither the eigenfunction nor the eigenvalue has
simple analytical expressions, and they are entangled with
the frequency eigenvalues of Fourier modes.
Although our ultimate goal is to overcome these difficulties altogether,
since each one is sufficiently involved, we choose to
proceed step by step, and tackle the subtraction problem first.

As a solution to the subtraction problem, we have recently
formulated two types of regularization methods \cite{MNS1},
extending the earlier work \cite{Mino,MinNak1}.
We termed the two methods the `power expansion regularization' and
the `mode-by-mode regularization'. Although the methods we have developed
should be applicable to the gravitational case in principle, we have
focused on the scalar case since it is free from the gauge problem:
The Regge-Wheeler or Teukolsky equation reduces to
the simple scalar d'Alembertian equation.
As for the mode-by-mode regularization, a variant of it
adapted to the gravitational case
has been applied to a particle in
circular orbit around a Schwarzschild black hole
in Ref.~\cite{NakSas} successfully but only to the 1PN order.
The purpose of the present paper is to demonstrate
the effectiveness of the power expansion regularization explicitly
and analytically by applying it to the simplest case of a scalar
particle in circular orbit around a Schwarzschild black hole
to the 3PN order.

Analytic solutions of the Teukolsky equation were derived by Mano,
Suzuki and Takasugi \cite{Mano,ManoTak} in the series of
hypergeometric functions. We use their result to construct
the Green function to the 3PN order where the technical advantage
of the power expansion regularization becomes clear.
The direct part has been obtained under the local expansion
for general orbits in Ref.~\cite{MNS1}.

There have appeared some papers discussing the extraction of the
 scalar self-force \cite{BarOri,Bur,BarBur,Bar,Lousto,BuLiSo}.
Recently, using the mode-sum regularization prescription
proposed in Ref.~\cite{BarOri},
the self-force acting on an electric or scalar charge has been
discussed in the spacetime of spherical shells.
Recently the nature of the gravitational self-force has been discussed by
Detweiler\cite{Detweil}.

This paper is organized as follows.
In Sec.~\ref{sec:MOR}, we briefly review the power expansion
regularization developed in Ref.~\cite{MNS1}.
In Sec.~\ref{sec:PEFSF},
we consider the full Green function to the 3PN order
in the Fourier-harmonic expanded form,
and perform the power expansion.
In Sec.~\ref{sec:PEDP}, we perform the power expansion of
the direct part evaluated by the local expansion.
Then in Sec.~\ref{sec:PRSRF},
we derive the regularized scalar self-force by
applying the power expansion regularization
to the results obtained in the previous two sections.
Finally, we summarize our result
and discuss possible extensions of our work
in Sec.~\ref{sec:CON}.

\section{Method of Regularization}\label{sec:MOR}

The reaction force on a point scalar charge is given by
\begin{eqnarray}
F^\alpha(\tau) = \lim_{x\rightarrow z(\tau)}
\left(F^\alpha[\phi^{\rm full}](x)-F^\alpha[\phi^{\rm dir}](x)\right)
=\lim_{x\to z(\tau)}F^\alpha[\phi^{\rm tail}](x)\,,
\end{eqnarray}
where
$\phi^{\rm full}$ is the scalar field induced by the point source,
$\phi^{\rm dir}$ is the direct part of the field defined
in Ref.~\cite{Qui,DeWBre,QuiWal,reaction},
$\phi^{\rm tail}=\phi^{\rm full}-\phi^{\rm dir}$,
and $F^\alpha[...]$ is a linear tensor derivative operator
to derive the force from the scalar field.
To obtain $\phi^{\rm full}$, we construct the full Green function
in the form of Fourier-harmonic series.
Since both the full scalar field and the direct part diverge
at the location of the particle,
we calculate them at a field point off the particle
where the subtraction of the divergent part is well-defined.
We call this field point $x_{\rm reg}^\alpha$ the regularization point.
There is freedom in the choice of the regularization point.
Let the location of the particle at its proper time $\tau=\tau_0$ be
\begin{eqnarray}
\{z^\alpha_{0}\} = \{z^\alpha(\tau_{0})\} = \{t_0,r_0,\theta_0,\phi_0\}\,.
\label{eq:ploc}
\end{eqnarray}
For the power expansion regularization, which will be
 briefly explained later in this section,
we take $\{x^\alpha_{\rm reg}\} = \{t,r,\theta_0,\phi_0\}$.

We consider a point scalar charge $q$ moving
in the Schwarzschild background,
\begin{eqnarray}
ds^2 = - \left(1-{2M \over r}\right)dt^2
+\left(1-{2M \over r}\right)^{-1}dr^2
 +r^2 \, (d\theta^2+\sin^2\theta d\phi^2) \,,
\end{eqnarray}
where $\{x^{\alpha}\}=\{t, r, \theta, \phi\}$ are the Schwarzschild
coordinates, and $M$ is the black hole mass.
The charge density of the scalar particle in circular orbit
is given by
\begin{eqnarray}
\rho(x^{\alpha})=q \int d \tau
{\delta^{(4)} (x^{\alpha}-z^{\alpha}(\tau)) \over \sqrt{-|g|} }
\label{eq:source} \,,
\end{eqnarray}
where $z^{\alpha}$ is the trajectory of the particle,
\begin{eqnarray}
z^\alpha(\tau)=
\left\{v^t \tau ,\, r_0 ,\, {\pi\over 2} ,\, v^\phi \tau \right\}
\,;\quad
v^t = \sqrt{r_0 \over r_0-3M} \,, \quad
v^\phi ~=~ {1\over r_0}\sqrt{M\over r_0-3M} \,,
\label{eq:orbit}
\end{eqnarray}
and $v^{\alpha}$ is the four velocity of the particle.
We have assumed
 the orbit is on the equatorial plane without loss of generality.

The full scalar field is calculated
by using the Green function method as
\begin{eqnarray}
\phi^{\rm full}(x) &=& q\int d\tau\, G^{\rm full}(x,z(\tau)) \,,
\label{eq:full-g}
\end{eqnarray}
where the Green function satisfies the wave equation
\begin{eqnarray}
\nabla^\alpha \nabla_\alpha G^{\rm full}(x,x') &=&
{\delta^{(4)}(x-x')\over \sqrt{-|g|}} \,,
\label{eq:wave}
\end{eqnarray}
and we impose the retarded boundary condition on $G^{\rm full}(x,x')$.
The full Green function is decomposed into
the Fourier-harmonic modes as
\begin{eqnarray}
G^{\rm full}(x,x') &=& \int {d\omega\over2\pi}{\omega\over|\omega|}
e^{-i\omega(t-t')}
\sum_{\ell m}
g^{\rm full}_{\ell m\omega}(r,r') Y_{\ell m}(\theta,\phi)Y^*_{\ell m}
(\theta',\phi')\,.
\label{eq:FHdeco-full-g}
\end{eqnarray}
Then the wave equation (\ref{eq:wave}) reduces to
an ordinary differential equation for the radial Green function as
\begin{eqnarray}
\left[\left(1-{2M\over r}\right){d^2\over dr^2}
+{2(r-M)\over r^2}{d\over dr}
+\left({\omega^2\over\displaystyle
 1-{2M\over r}}-{\ell(\ell+1)\over r^2}\right)
\right]g^{\rm full}_{\ell m\omega}(r,r')
&=& {1\over r^2}\delta(r-r') \,. \label{eq:radial-eq}
\end{eqnarray}
We construct the radial function of the full Green function
by using homogeneous solutions of (\ref{eq:radial-eq})
which can be obtained by a systematic analytic method
developed in Ref.~\cite{Mano}.

On the other hand, the local expansion of the full Green function
gives the direct part of the scalar field,
\begin{eqnarray}
\phi^{\rm dir}(x) &=& q\int d\tau\, G^{\rm dir}(x,z(\tau)) \,,
\end{eqnarray}
where $G^{\rm dir}$ is the direct part of the Green function.
It is given in a covariant manner as
\begin{eqnarray}
G^{\rm dir}(x,x') &=& -{1\over 4\pi}\theta[\Sigma(x),x']
\sqrt{\Delta(x,x')}\delta\bigl(\sigma(x,x')\bigr) \,,
\label{eq:direct-g}
\end{eqnarray}
where $\sigma(x,x')$ is the bi-scalar of half the squared geodesic distance,
$\Delta(x,x')$ is the generalized van Vleck-Morette determinant bi-scalar,
$\Sigma(x)$ is an arbitrary spacelike hypersurface containing $x$,
and $\theta[\Sigma(x),x']=1-\theta[x',\Sigma(x)]$ is equal to 1
when $x'$ lies in the past of $\Sigma(x)$
and vanishes when $x'$ lie in the future.

We now briefly describe the power expansion regularization procedure.
As mentioned earlier, for a given point on
 the particle trajectory (\ref{eq:ploc}),
we choose the regularization point as
\begin{eqnarray}
\{x^\alpha_{\rm reg}\} = \{t,r,\theta_0,\phi_0\} \,,
\end{eqnarray}
which is assumed to be on a suitably chosen spacelike hypersurface:
See Eq.~(3.8) in Ref.~\cite{MNS1}.
It should be noted that $t$ is a function of $r$.
In the following, we take $\tau_0=0$ for simplicity.

Assuming $(r-r_0)/r_0\ll1$, we expand the full scalar field
and its direct part in powers of the radius $r$ as
\begin{eqnarray}
\phi^{\rm full}(x_{\rm reg}) &=& \sum_n r^n \phi^{{\rm full}(n)} \,,
\quad
\phi^{\rm dir}(x_{\rm reg})=\sum_n r^n \phi^{{\rm dir}(n)} \,,
\end{eqnarray}
Similarly, for the calculation of the self-force, we expand
$\nabla_\mu\phi^{{\rm full}}$ and $\nabla_\mu\phi^{{\rm dir}}$
as
\begin{eqnarray}
\nabla_\mu\phi^{{\rm full}}(x_{\rm reg})
&=&\sum_n r^n \phi_\mu^{{\rm full}(n)}\,,
\quad
\nabla_\mu\phi^{{\rm dir}}(x_{\rm reg})
=\sum_n r^n \phi_\mu^{{\rm dir}(n)}\,.
\end{eqnarray}
For $\phi^{\rm full}$ and $\nabla_\mu\phi^{\rm full}$,
this expansion is done first for each spherical
harmonic mode, followed by summation of the Fourier-harmonic components
to obtain $\phi^{{\rm full}(n)}$ and $\phi_\mu^{{\rm full}(n)}$.
The heart of the power expansion regularization resides in the fact that
the coefficients $\phi^{{\rm full}(n)}$ and $\phi_\mu^{{\rm full}(n)}$
as well as $\phi^{{\rm dir}(n)}$ and $\phi_\mu^{{\rm dir}(n)}$
of the power series are finite,
though these series as a whole diverge in the limit $r\to r_0$.
Hence we may extract the tail part as
\begin{eqnarray}
\phi^{\rm tail}(x_{\rm reg}) &=& \sum_n r^n \phi^{{\rm tail}(n)} \,;
\quad
\phi^{{\rm tail}(n)} = \phi^{{\rm full}(n)}-\phi^{{\rm dir}(n)} \,.
\nonumber\\
\nabla_\mu\phi^{\rm tail}(x_{\rm reg})
 &=& \sum_n r^n \phi_\mu^{{\rm tail}(n)} \,;
\quad
\phi_\mu^{{\rm tail}(n)}
= \phi_\mu^{{\rm full}(n)}-\phi_\mu^{{\rm dir}(n)} \,.
\end{eqnarray}
The tail part of the scalar field is obtained by summing over $n$
and taking the limit $r\to r_0$.
This extraction procedure is called the power expansion regularization.

\section{Power Expansion of the Full Scalar Field}\label{sec:PEFSF}

We first consider the full Green function
decomposed in the Fourier-harmonic modes (\ref{eq:FHdeco-full-g}),
The retarded radial Green function becomes
\begin{eqnarray}
&&g_{\ell m\omega}^{\rm full}(r,r')
= {1\over W_{\ell m\omega}(\phi^{\rm in},\phi^{\rm up})}
\left(\phi_{\ell m\omega}^{\rm in}(r)
\phi_{\ell m\omega}^{\rm up}(r')\theta(r'-r)
+\phi_{\ell m\omega}^{\rm up}(r)
\phi_{\ell m\omega}^{\rm in}(r')\theta(r-r')\right) \,;
\nonumber\\
&&\quad
W_{\ell m\omega}(\phi^{\rm in},\phi^{\rm up})
= r^2\left(1-{2M\over r}\right)
\left(\biggl({d\over dr}\phi_{\ell m\omega}^{\rm in}(r)\biggr)
\phi_{\ell m\omega}^{\rm up}(r)
-\biggl({d\over dr}\phi_{\ell m\omega}^{\rm up}(r)\biggr)
\phi_{\ell m\omega}^{\rm in}(r)\right) \,,
\label{eq:fullG}
\end{eqnarray}
where $\phi_{\ell m\omega}^{\rm in}$ is a homogeneous solution
which vanishes on the past horizon (when multiplied by
$e^{-i\omega t}$),
and $\phi_{\ell m\omega}^{\rm up}$ is a homogeneous solution
which vanishes at the past null infinity \cite{ChrMis}.
They are called the in-going and up-going solutions, respectively.

Homogeneous solutions of (\ref{eq:radial-eq})
were fully investigated in Ref.~\cite{Mano,ManoTak}
and we have various analytic expressions for the homogeneous functions.
In this paper, we consider the slow-motion expansion, in which we assume
\begin{eqnarray}
\omega r ~\approx~ v \,,\quad
\omega M ~\approx~ v^3 \,;\quad v ~\ll~ 1 \,.
\end{eqnarray}
Detailed properties of the radial Green function
for the calculation up to the 3PN order are
described in Appendix \ref{app:GF}.
For technical reasons,
we classify the harmonic components of the Green function
into the three parts,
\begin{eqnarray}
G^{{\rm (sym)}}_{(\ell \geq 2)}(x,x')
&=& \int{d\omega\over 2\pi}{\omega\over|\omega|}e^{-i\omega(t-t')}
 \sum_{\ell \geq 2,m}
g^{\rm (sym)}_{\ell m\omega}(r,r')Y_{\ell m}(\theta,\phi)
Y^*_{\ell m}(\theta',\phi')
\,, \\
G^{{\rm (rad)}}_{(\ell \geq 2)}(x,x')
 &=& \int{d\omega\over 2\pi}{\omega\over|\omega|}e^{-i\omega(t-t')}
 \sum_{\ell \geq 2,m}
g^{\rm (rad)}_{\ell m\omega}(r,r')Y_{\ell m}(\theta,\phi)
Y^*_{\ell m}(\theta',\phi')
\,, \\
G_{(\ell=0,1)}^{\rm full}(x,x')
&=& \int{d\omega\over 2\pi}{\omega\over|\omega|}e^{-i\omega(t-t')}
 \sum_{\ell=0,1,m}
g_{\ell m\omega}^{\rm full}(r,r')Y_{\ell m}(\theta,\phi)
Y^*_{\ell m}(\theta',\phi')
\,.
\end{eqnarray}
The suffices (sym) and (rad) represent that they are
the time-symmetric and radiative parts, respectively, of the
Green function. That is, we write the (retarded) Green function as
\begin{eqnarray}
G^{\rm full}&=&G^{\rm(sym)}+G^{\rm(rad)}\,;
\nonumber\\
&&G^{\rm(sym)}={1\over2}(G^{\rm(ret)}+G^{\rm(adv)})\,,
\quad
G^{\rm(rad)}={1\over2}(G^{\rm(ret)}-G^{\rm(adv)})\,,
\end{eqnarray}
where $G^{\rm(adv)}$ stands for the advanced Green function.
Correspondingly, the full scalar field is
divided into the three parts as
\begin{eqnarray}
\phi^{\rm (sym)}_{(\ell \geq 2)}(x)
&=& \int d^4x'\sqrt{-|g|(x')}G^{\rm (sym)}_{(\ell \geq 2)}(x,x') \rho(x')\,,
\label{eq:phi-inhomo}\\
\phi^{\rm (rad)}_{(\ell \geq 2)}(x)
&=& \int d^4x'\sqrt{-|g|(x')}G^{\rm (rad)}_{(\ell \geq 2)}(x,x') \rho(x')\,,
\label{eq:phi-homo}\\
\phi_{(\ell=0,1)}^{\rm full}(x)
&=& \int d^4x'\sqrt{-|g|(x')}G_{(\ell=0,1)}^{\rm full}(x,x') \rho(x')\,.
\label{eq:phi-01}
\end{eqnarray}

The division of the Green function to the harmonic modes
of $\ell\geq2$ and $\ell=0,1$ is purely technical
because of some non-systematic analytical behavior of the modes
$\ell=0,1$ as discussed in Appendix~\ref{app:GF}.
On the other hand, there is a physical reason
for dividing the Green function to
the time-symmetric part and the radiative part.
It is known that the radiation reaction to
the energy and angular momentum of the particle
(i.e., any conserved quantity associated with a background
 Killing vector field) can be calculated
by using the radiative Green function $G^{\rm(rad)}$ \cite{Galtsov}.
In our case of circular orbit around a Schwarzschild black hole,
we therefore expect no contribution from the time-symmetric part
to the $t$- and $\phi$-components of the reaction force.
Nevertheless, one cannot simply discard the
time-symmetric part since the tail part is not identical to
the radiative part. In particular, the $r$-component of the
reaction force, which does not contribute to the energy and
angular momentum loss in our case, cannot be obtained
from the radiative part. Since our purpose here is to demonstrate
 the validity of the power expansion regularization, we regularize
$G^{{\rm (sym)}}_{(\ell \geq 2)}$ to show the vanishing of
the $t$- and $\phi$-components of the reaction force
due to the time-symmetric part explicitly,
as well as to derive the non-vanishing $r$-component of
the reaction force.
As for $G^{{\rm (rad)}}_{(\ell \geq 2)}$ and $G_{(\ell=0,1)}^{\rm full}$,
they are always finite. Hence the power expansion is unnecessary
for these parts.

The time-symmetric part of the radial Green function
is expanded in powers of $r$ as
\begin{eqnarray}
g^{\rm (sym)}_{\ell m\omega}(r,r') &=& \sum_n
\left(\theta(r'-r){r^{n+\ell}\over r'^{n+\ell+1}}
g^{{\rm in}(n)}_{\ell m\omega}(r')
+\theta(r-r'){r'^{n+\ell}\over r^{n+\ell+1}}
g^{{\rm out}(n)}_{\ell m\omega}(r')
\right) \,.
\end{eqnarray}
We note that because of the singularity at $r=r_0$,
we have two different forms of the power expansion at $0<r<r_0$ and $r_0<r$.
The final regularized force, however, is the same for either
choice of the regularization point.

For the calculation to the 3PN order,
it is found that we only need $g^{{\rm in}(n)}_{\ell m\omega}$
for $-3\le n\le 6$ and $g^{{\rm out}(n)}_{\ell m\omega}$
for $-6\le n\le 3$.
The coefficients $g^{{\rm in}(n)}_{\ell m\omega}$
and $g^{{\rm out}(n)}_{\ell m\omega}$
are summarized in Appendix~\ref{app:coef}.
After summing over $\omega$, $\ell$ and $m$,
we obtain the power expansion of the time-symmetric Green function as
\begin{eqnarray}
G^{\rm (sym)}_{(\ell \geq 2)}(x,x')
= \int{d\omega\over 2\pi}{\omega\over|\omega|}e^{-i\omega(t-t')}
\sum_n
&&\Biggl(\theta(r'-r){r^n\over r'^{n+1}}
G^{{\rm in}(n)}_{(\ell \geq 2)}(\theta,\phi;r',\theta',\phi';\omega)
\nonumber \\
&&+\theta(r-r'){r'^n\over r^{n+1}}
G^{{\rm out}(n)}_{(\ell \geq 2)}(\theta,\phi;r',\theta',\phi';\omega)\Biggr),
\end{eqnarray}
where
\begin{eqnarray}
G^{{\rm in/out}(n)}_{(\ell \geq 2)}(\theta,\phi;r',\theta',\phi';\omega)
=\sum_{\ell \geq 2,m} g_{\ell m\omega}^{{\rm in/out}(n-\ell)}(r')
Y_{\ell m}(\theta,\phi)Y^*_{\ell m}(\theta',\phi') \,.
\end{eqnarray}
One sees that only a finite number of harmonic modes contribute
to each expansion coefficient $G^{{\rm in/out}(n)}$
because of the truncation of the terms at large $|n-\ell|$,
which greatly eases the calculation.

Now the time-symmetric part of the scalar field (\ref{eq:phi-inhomo})
becomes
\begin{eqnarray}
\phi^{\rm (sym)}_{(\ell \geq 2)}(x)
&=& {q\over v^t} \sum_n
\Biggl(\theta(r_0-r){r^n\over r_0^{n+1}}\phi^{{\rm in}(n)}(r_0)
+\theta(r-r_0){r_0^n\over r^{n+1}}\phi^{{\rm out}(n)}(r_0)\Biggr)
\,,
\end{eqnarray}
where
\begin{eqnarray}
\phi^{{\rm in/out}(n)} =
\sum_{\ell m} g_{\ell m,m\Omega}^{{\rm in/out}(n-\ell)}(r_0)
Y_{\ell m}(\theta,\phi)
Y^*_{\ell m}(\pi/2,0) e^{-im\Omega t} \,,
\end{eqnarray}
and we have introduced the angular frequency $\Omega = v^\phi/v^t$.
Taking the derivative of the scalar field,
we have the power expanded scalar field at the regularization point
as
\begin{eqnarray}
\nabla_t \phi^{\rm (sym)}_{(\ell \geq 2)}(x_{\rm reg})
&=&0
\label{eq:full-power-t} \,,
 \\
\nabla_r \phi^{\rm (sym)}_{(\ell \geq 2)}(x_{\rm reg})
&=& {q\over v^t} \sum_n
\Biggl(\theta(r_0-r)n{r^{n-1}\over r_0^{n+1}}\phi^{{\rm in}(n)}_{(r)}(r_0)
-\theta(r-r_0)(n+1){r_0^n\over r^{n+2}}\phi^{{\rm out}(n)}_{(r)}(r_0)\Biggr)
\label{eq:full-power-r} \,, \\
\nabla_\theta \phi^{\rm (sym)}_{(\ell\geq2)}(x_{\rm reg})
&=& 0
\label{eq:full-power-theta} \,, \\
\nabla_\phi \phi^{\rm (sym)}_{(\ell\geq2)}(x_{\rm reg})
&=&
 -{1\over \Omega} \nabla_t \phi^{\rm (sym)}_{(\ell\geq2)}(x_{\rm reg})
=0\,,
\label{eq:full-power-phi}
\end{eqnarray}
where
\begin{eqnarray}
\phi^{{\rm in/out}(n)}_{(r)} &=&
\sum_{\ell m} g_{\ell m,m\Omega}^{{\rm in/out}(n-\ell)}(r_0)
|Y_{\ell m}(\pi/2,0)|^2
\label{eq:full-power-r0} \,.
\end{eqnarray}
The fact that the $\theta$-component vanishes comes from
our assumption of circular orbit, while the
 reason for the vanishing $t$- and  $\phi$-components is a bit subtle.
It comes both from the assumption of circular orbit and
the symmetry of $g_{\ell m\omega}^{\rm(sym)}$
 under $\omega\leftrightarrow-\omega$.
The formula used for the $m$-sum
in Eq.~(\ref{eq:full-power-r0})
is presented in Appendix \ref{app:m-sum}.

\section{Power Expansion of the Direct Part}\label{sec:PEDP}

Here we evaluate the derivative of the direct part of the scalar field
under the local expansion.
The details of this evaluation procedure are discussed in Ref.~\cite{MNS1}.
By calculating Eqs.~(3.15-7) in Ref.~\cite{MNS1},
we obtain the regularized direct part of the scalar reaction
force as
\begin{eqnarray}
\nabla_t \phi^{\rm dir}(x_{\rm reg}) &=& O(r-r_0) \,, \\
\nabla_r \phi^{\rm dir}(x_{\rm reg}) &=&
{q\over 4\pi}\sqrt{1-{2M\over r_0}}\,{{\rm sgn}(r-r_0)\over (r-r_0)^2}
\left(
1-{1\over 4}{M^2(r_0-M)\over r_0^2(r_0-2M)^2(r_0-3M)}(r-r_0)^2
\right) \nonumber \\
&& \qquad \qquad
+O(r-r_0) \,,
\label{eq:dir-force-r}\\
\nabla_\theta \phi^{\rm dir}(x_{\rm reg}) &=& O(r-r_0) \,, \\
\nabla_\phi \phi^{\rm dir}(x_{\rm reg}) &=& O(r-r_0) \,,
\end{eqnarray}
where ${\rm sgn}(r-r_0)=\pm1$ for $r-r_0\gtrless0$, and
we have ignored the terms of order $r-r_0$
as they do not contribute to the force in the
 coincidence limit $r\rightarrow r_0$.
It should be mentioned that the above are exact in the sense that
no slow-motion approximation is employed.
Note that the second term proportional to $M^2/r_0^2$ in the
parentheses on the right-hand-side of the radial component
gives a manifestly finite contribution in the limit $r\to r_0$.
Of course, these simple forms of the direct force components
owe to the assumption of circular orbit: They will be
more complicated if the radial velocity is non-zero.

Using the formula,
\begin{eqnarray}
{1\over (r-r_0)^2} &=& \sum_{n\geq0}
\left(\theta(r-r_0)(n+1){r_0^n\over r^{n+2}}
+\theta(r_0-r)(n+1){r^n\over r_0^{n+2}}\right) \,,
\end{eqnarray}
we finally obtain the power expansion of the direct part as
\begin{eqnarray}
\nabla_t \phi^{\rm dir}(x_{\rm reg}) &=& 0
\label{eq:dir-power-t} \,, \\
\nabla_r \phi^{\rm dir}(x_{\rm reg})
&=& {q\over 4\pi}\sqrt{1-{2M\over r_0}}
\Biggl\{\theta(r-r_0)\left(\sum_{n\geq0}(n+1){r_0^n\over r^{n+2}}
-{1\over 4}{M^2(r_0-M)\over r_0^2(r_0-2M)^2(r_0-3M)}\right)
\nonumber \\ && \qquad \qquad \qquad
-\theta(r_0-r)\left(\sum_{n\geq0}(n+1){r^n\over r_0^{n+2}}
-{1\over 4}{M^2(r_0-M)\over r_0^2(r_0-2M)^2(r_0-3M)}\right)
\Biggr\} \label{eq:dir-power-r} \,, \\
\nabla_\theta \phi^{\rm dir}(x_{\rm reg}) &=& 0 \,, \\
\nabla_\phi \phi^{\rm dir}(x_{\rm reg}) &=& 0
\label{eq:dir-power-phi} \,.
\end{eqnarray}
We find that only the $r$-component of the force diverges in the
limit $r\to r_0$.

\section{Extraction of the Scalar Reaction Force}\label{sec:PRSRF}

Now we are ready to discuss the extraction of the tail part
under the power expansion regularization.
Since the $r$-component of the full (bare) force
is the only one that diverges in the coincidence limit,
we first consider the simpler cases of the
$t$-, $\theta$- and $\phi$-components, and separately treat
the $r$-component afterwards.

For the $\theta$-component, all the parts of the force vanish
identically because of the assumption of circular orbit,
\begin{eqnarray}
\nabla_\theta \phi^{\rm tail}(z) =0\,.
\end{eqnarray}
For the $t$- and $\phi$-components, which are equal to each other
modulo the factor $-1/\Omega$, both the time-symmetric part
and the direct part of the force vanish identically,
as given in Eqs.~(\ref{eq:full-power-t}) and (\ref{eq:dir-power-phi}).
Hence the only possible contributions come from
$\nabla_t\phi_{(\ell\geq2)}^{\rm(rad)}$
and $\nabla_t\phi_{(\ell=0,1)}$, in accordance with
our expectation.
We find the former contribution arises at the 3PN order and
the latter at the 2PN order, which are added together to give
\begin{eqnarray}
\nabla_t \phi^{\rm tail}(z)
 &=&\nabla_t \phi^{\rm (rad)}_{(\ell \geq 2)}(z)
+\nabla_t \phi_{(\ell=0,1)}(z)
\nonumber \\
&=& {q \over 4\pi r_0^2}\left({1 \over 3}v^4+{46073 \over 30}v^6
+O(v^7) \right)  \,, \\
\nabla_\phi \phi^{\rm tail}(z)
&=& -{1\over \Omega} \nabla_t \phi^{\rm tail}(z)
\nonumber \\
&=& -{q \over 4\pi r_0}\left({1 \over 3}v^3+{46073 \over 30}v^5
+O(v^7) \right)
\,.
\end{eqnarray}
It is easy to check that the leading terms of the above are
consistent with that of the energy loss rate by scalar radiation
evaluated from the energy-momentum tensor
of the scalar field at future null-infinity. The leading terms
are also in agreement with Gal'tsov \cite{Galtsov}.

Now we turn to the $r$-component of the reaction force.
Since only the subtraction of the divergent terms is non-trivial,
taking account of the form of $\nabla_r \phi^{\rm dir}(x_{\rm reg})$
given by Eq.~(\ref{eq:dir-force-r}),
we introduce the auxiliary coefficients $f^{{\rm in/out}(n)}$
in the extraction procedure of the tail part from
$\nabla_r \phi_{(\ell\geq2)}^{\rm (sym)}(x_{\rm reg})$ as
\begin{eqnarray}
&& \nabla_r \phi^{\rm (sym)}_{(\ell \geq 2)}(x_{\rm reg})
-{q\over 4\pi}\sqrt{1-{2M\over r_0}}{{\rm sgn}(r-r_0)\over (r-r_0)^2}
\nonumber \\ && \qquad \qquad
~=~ {q\over 4\pi} \sum_n
\Biggl(\theta(r_0-r)n{r^{n-1}\over r_0^{n+1}}f^{{\rm in}(n)}
-\theta(r-r_0)(n+1){r_0^n\over r^{n+2}}f^{{\rm out}(n)}\Biggr) \,.
\label{eq:force-sum}
\end{eqnarray}
Then we obtain
\begin{eqnarray}
f^{{\rm in}(n)} &=& O(v^7) \qquad (n<1)
\,, \\
f^{{\rm in}(1)} &=& 1+v^2+{11\over 10}v^4+{331\over 100}v^6
+O(v^7) \,, \\
f^{{\rm in}(2)} &=& -{33\over 20}v^4-{173713\over 14700}v^6
+O(v^7) \,, \\
f^{{\rm in}(3)} &=& -{3\over 10}v^2-{3\over 10}v^4+{215419\over 22050}v^6
+O(v^7) \,, \\
f^{{\rm in}(4)} &=& -{30154441\over 9604980}v^6
+O(v^7) \,, \\
f^{{\rm in}(5)} &=& {1\over 56}v^4+{237035\over 30918888}v^6
+O(v^7) \,, \\
f^{{\rm in}(6)} &=& -{44372\over 10735725}v^6
+O(v^7) \,, \\
f^{{\rm in}(7)} &=& -{1288333\over 527482800}v^6
+O(v^7) \,, \\
f^{{\rm in}(n)} &=&
{-6n(939-912n-224n^3+224n^4-1972n^2)\over
(2n-7)(2n-5)(2n-3)(2n-1)^2(2n+1)^2(2+3)^2} v^6
+O(v^7) \qquad (n>7) \,, \\
f^{{\rm out}(n)} &=& O(v^7) \qquad (n<-4)
\,, \\
f^{{\rm out}(-4)} &=& v^6
+O(v^7) \,, \\
f^{{\rm out}(-3)} &=& -{19\over 15}v^6
+O(v^7) \,, \\
f^{{\rm out}(-2)} &=& -{1\over 2}v^4+{13\over 9}v^6
+O(v^7) \,, \\
f^{{\rm out}(-1)} &=& +O(v^7)
\,, \\
f^{{\rm out}(0)} &=& -1+{3\over 2}v^2-{7\over 8}v^4-{7079\over 5040}v^6
+O(v^7) \,, \\
f^{{\rm out}(1)} &=& -2+{16\over 5}v^2+{117\over 35}v^4-{58711\over 69300}v^6
+O(v^7) \,, \\
f^{{\rm out}(2)} &=& -6v^2+{58\over 5}v^4+{3686944\over 525525}v^6
+O(v^7) \,, \\
f^{{\rm out}(3)} &=& -{72\over 5}v^4+{12656296\over 429975}v^6
+O(v^7) \,, \\
f^{{\rm out}(4)} &=& -{3396569326\over 106135029}v^6
+O(v^7) \,, \\
f^{{\rm out}(n)} &=&
{-6(n+1)(327+1120n^3+224n^4-1464n+44n^2)\over
(2n-1)^2(2n+1)^2(2n+3)^2(2n+5)(2n+7)(2n+9)}
+O(v^7) \qquad (n>4) \,,
\end{eqnarray}
where $v=\Omega r_0$ (and $\Omega M = v^3$).

As for the parts $\nabla_r\phi^{\rm (rad)}_{(\ell \geq 2)}$ and
$\nabla_r\phi_{(\ell=1,0)}$,
we find the former is higher than the 3PN order and
only the latter contributes at the 3PN order.
In fact, the radiative Green function
does not contribute to the radial force
for a circular orbit to full order (see Appendix~\ref{app:GF}).

By summing up (\ref{eq:force-sum})
and adding the finite terms in the direct part together with
$\nabla_r\phi_{(\ell=1,0)}$,
we finally find
\begin{eqnarray}
\nabla_r \phi^{\rm tail}(z)
&=& \lim_{r\rightarrow r_0\pm0}\left(
\nabla_r \phi^{\rm (sym)}(x_{\rm reg})
-\nabla_r \phi^{\rm dir}(x_{\rm reg})\right)
+\nabla_r \phi^{\rm (rad)}_{(\ell \geq 2)}(z)
+\nabla_r \phi_{(\ell=0,1)}(z)
\nonumber \\
&=& -{q\over 4\pi r_0^2}\left({31\over 12}v^6-{7\over 64}\pi^2 v^6
+{8\over 3}v^6\ln (v) +O(v^7) \right) \,.
\end{eqnarray}
As expected, both choices of the regularization point,
either at $r<r_0$ or $r>r_0$, give the same reaction force
in the limit $r\to r_0$. 
It is noted that the above result is consistent with 
that of the numerical calculation in Ref.~\cite{Bur}. 

\section{Conclusion}\label{sec:CON}

In a separate paper\cite{MNS1}, we have proposed two technical methods to
derive the scalar reaction force on a particle in the
Schwarzschild background, the power expansion regularization
and the mode-by-mode regularization. We expect these methods
will be useful also for the case of gravitational reaction
force. In fact, a variant of the mode-by-mode regularization method
has been successfully applied to the gravitational case, although
only the 1PN calculation for circular orbit has been done \cite{NakSas}.

In this paper, we have demonstrated the usefulness of the other
regularization method, the power expansion regularization method,
by applying it to a scalar particle
in circular orbit around a Schwarzschild black hole
and calculating the scalar reaction force to the 3PN order.
In the power expansion regularization method,
the tail part of the scalar field, which is responsible for the reaction
force, is obtained by subtracting the direct part
from the full scalar field at a regularization point specified by
the radial coordinate $r$, followed by
taking the limit $r\to r_0$ where $r_0$ is the
radial coordinate of the particle.
We have found that the $r$-component of the scalar reaction force arises
at the 3PN order, and the $t$- and $\phi$-components
at the 2PN and 1.5PN orders, respectively.

Since the derivation of reaction force for
non-circular orbits is very important for binary systems of
extreme mass ratio, and the gravitational waves from such
binaries are expected to be detected by
future space-based gravitational wave detectors such as LISA,
it is indispensable for us to proceed to the case of
general orbit.
In this respect, although the calculation is bound to become more complicated
when we consider non-circular orbits where the radial motion
is present, the power expansion regularization method seems
to be still applicable. A general framework for general orbit
in terms of the power expansion regularization scheme has
been developed in Ref.~\cite{MNS1}.
As a next step, we plan to consider the eccentric or the radial orbit
with these regularization methods. 


\noindent

\section*{Acknowledgements}

We thank U.~Gen, Y.~Himemoto and F.~Takahara for fruitful conversations.
Special thanks are due to H.~Tagoshi and T.~Tanaka for valuable discussions.
HN and YM also thank K.S.~Thorne for his hospitality and discussions
during their stay at CalTech, and
YM thanks B.~Schutz for his hospitality and discussions
during his stay at the Einstein Institute in Potsdam.
This work was supported in part by Monbukagakusho Grant-in-Aid
for Creative Research, No.~09NP0801, and by
Monbukagakusho Grant-in-Aid for Scientific Research, No.~12640269.
HN and YM are supported by Research Fellowships of the
Japan Society for the Promotion of Science
for Young Scientists, No.~2397 and No.~0704.


\begin{appendix}

\section{Full Green Function} \label{app:GF}

In this appendix, we summarize the derivation of the
full radial Green function.
Here the in-going and up-going
homogeneous solutions are denoted by $\phi_{\rm in}^{\nu}$ and
$\phi_{\rm up}^{\nu}$, respectively, where $\nu$ is called the
renormalized angular momentum \cite{Mano,ManoTak},
and is equal to $\ell$ in the limit $\omega M\to0$.

The radial Green function is expressed as
\begin{eqnarray}
&&g_{\ell m\omega}^{\rm full}(r,r')
= {1\over W_{\ell m\omega}(\phi_{\rm in}^{\nu},\phi_{\rm up}^{\nu})}
\left(\phi_{\rm in}^{\nu}(r)\phi_{\rm up}^{\nu}(r')\theta(r'-r)
+\phi_{\rm up}^{\nu}(r)\phi_{\rm in}^{\nu}(r')\theta(r-r')\right) \,;
\nonumber\\
&&\quad
W_{\ell m\omega}(\phi_{\rm in}^{\nu},\phi_{\rm up}^{\nu})
= r^2\left(1-{2M\over r}\right)
\left(\biggl({d\over dr}\phi_{\rm in }^{\nu}(r)\biggr)
\phi_{\rm up}^{\nu}(r)
-\biggl({d\over dr}\phi_{\rm up}^{\nu}(r)\biggr)
\phi_{\rm in}^{\nu}(r)\right) \,.
\end{eqnarray}
For technical convenience, we express the homogeneous solutions
$\phi_{\rm in}^\nu$ and $\phi_{\rm up}^\nu$ in terms of
the solutions $\phi_{\rm c}^{\nu}$ and $\phi_{\rm c}^{-\nu-1}$,
which are expressed in terms of a series of the Coulomb wave
functions \cite{Mano}, as
\begin{eqnarray}
\phi_{\rm in}^{\nu} &=&
\alpha_\nu\phi_{\rm c}^{\nu}+\beta_{\nu}\,\phi_{\rm c}^{-\nu-1} \,,
\nonumber\\
\phi_{\rm up}^{\nu} &=&
\gamma_{\nu}\,\phi_{\rm c}^{\nu}+\delta_\nu\phi_{\rm c}^{-\nu-1} \,,
\end{eqnarray}
where the properties and the relations of the coefficients
$\{\alpha_{\nu},\,\beta_{\nu},\,\gamma_{\nu},\,\delta_{\nu}\}$
have been discussed extensively in Ref.~\cite{Mano,ManoTak}.
The function $\phi_{\rm c}^\nu$ is denoted by $R_c^\nu$ in
Ref.~\cite{Mano}.
It may be noted that $\phi_{\rm c}^\nu(r)\propto j_\ell(\omega r)$
in the limit $\epsilon\to0$, where $j_\ell$ is the spherical Bessel
function, and we may choose its phase so that $\phi_{\rm c}^\nu$ is
real.

We then divide the Green function to the time-symmetric and radiative parts,
\begin{eqnarray}
&&g_{\ell m\omega}^{\rm full}(r,r')=
g_{\ell m\omega}^{\rm (sym)}(r,r') +g_{\ell m\omega}^{\rm (rad)}(r,r') \,,
\end{eqnarray}
where
\begin{eqnarray}
g_{\ell m\omega}^{\rm (sym)}(r,r') &&=
{1\over W_{\ell m\omega}(\phi_{\rm c}^{\nu},\phi_{\rm c}^{-\nu-1})}
\nonumber\\
\times&&\Biggl[
\theta(r'-r)\Biggl\{
\phi_{\rm c}^{\nu}(r)\phi_{\rm c}^{-\nu-1}(r')
+\Re\left[{\tilde\beta_\nu\tilde\gamma_\nu
\over1-\tilde\beta_\nu\tilde\gamma_\nu}\right]
\left(\phi_{\rm c}^{\nu}(r)\phi_{\rm c}^{-\nu-1}(r')
+\phi_{\rm c}^{-\nu-1}(r)\phi_{\rm c}^{\nu}(r')\right)
\nonumber\\
&&\qquad\qquad
+\Re\left[{\tilde\gamma_\nu\over1-\tilde\beta_\nu\tilde\gamma_\nu}\right]
\phi_{\rm c}^{\nu}(r)\phi_{\rm c}^{\nu}(r')
+\Re\left[{\tilde\beta_\nu\over1-\tilde\beta_\nu\tilde\gamma_\nu}\right]
\phi_{\rm c}^{-\nu-1}(r)\phi_{\rm c}^{-\nu-1}(r')\Biggr\}
\nonumber\\
&&\qquad+\theta(r-r')\Biggl\{\nu\leftrightarrow-\nu-1\,,
\quad \tilde\beta_\nu\leftrightarrow\tilde\gamma_\nu\Biggr\}
\Biggr]\,,
\label{eq:gsym} \\
g_{\ell m\omega}^{\rm (rad)}(r,r') &&=
{i\over W_{\ell m\omega}(\phi_{\rm c}^{\nu},\phi_{\rm c}^{-\nu-1})}
\Biggl[\Im\left[{\tilde\beta_\nu\tilde\gamma_\nu
     \over1-\tilde\beta_\nu\tilde\gamma_\nu}\right]
\left(\phi_{\rm c}^{\nu}(r)\phi_{\rm c}^{-\nu-1}(r')
+\phi_{\rm c}^{-\nu-1}(r)\phi_{\rm c}^{\nu}(r')\right)
\nonumber\\
&&\qquad\qquad\qquad\qquad
+\Im\left[{\tilde\gamma_\nu\over1-\tilde\beta_\nu\tilde\gamma_\nu}\right]
\phi_{\rm c}^{\nu}(r)\phi_{\rm c}^{\nu}(r')
+\Im\left[{\tilde\beta_\nu\over1-\tilde\beta_\nu\tilde\gamma_\nu}\right]
\phi_{\rm c}^{-\nu-1}(r)\phi_{\rm c}^{-\nu-1}(r')
\Biggr]\,.
\label{eq:grad}
\end{eqnarray}
In the above, we have assumed $\alpha_\nu\neq0$ and $\delta_\nu\neq0$,
and introduced the coefficients $\{\tilde\beta_\nu,\,\tilde\gamma_\nu\}
:=\{\beta_\nu/\alpha_\nu,\,\gamma_\nu/\delta_\nu\}$.
Using the result obtained in Ref.~\cite{Mano},
we find the coefficients $\{\tilde\beta_\nu,\,\tilde\gamma_\nu\}$
behave under post-Newtonian expansion as
\begin{eqnarray}
\Re[\tilde\beta_\nu]=O(v^{6\ell+3})\,,
\quad
\Im[\tilde\beta_\nu]=O(v^{6\ell})\,,
\quad
\Re[\tilde\gamma_\nu]=O(v^6)\,,
\quad
\Im[\tilde\gamma_\nu]=(-1)^{\ell+1}\,.
\label{eq:betagamma}
\end{eqnarray}

Although the analysis in Ref.~\cite{Mano,ManoTak} is
very useful, it turns out that
the post Newtonian expansion of $\phi_c^{-\nu-1}$
breaks down at the $(\ell+2)$th PN order.
Thus for our analysis to the 3PN order,
the modes $\ell=0,1$ need a separate treatment.
So, we consider the three cases,
$g^{\rm(sym)}_{\ell m\omega}$ and $g^{\rm(rad)}_{\ell m\omega}$
for $\ell\geq2$, and $g_{\ell m\omega}$ for $\ell=0,1$, separately.

\subsection{Time-symmetric part for $\ell \geq 2$}\label{app:L2I}

To the 3PN order, inspection of the post-Newtonian orders of
$\{\tilde\beta_\nu,\,\tilde\gamma_\nu\}$ shows that
the time-symmetric Green function for $\ell\geq2$
can be approximated as
\begin{eqnarray}
g_{\ell m\omega}^{\rm (sym)}(r,r') =
{1\over W_{\ell m\omega}(\phi_{\rm c}^{\nu},\phi_{\rm c}^{-\nu-1})}
\left[
\theta(r'-r)\phi_{\rm c}^{\nu}(r)\phi_{\rm c}^{-\nu-1}(r')
+\theta(r-r')\phi_{\rm c}^{-\nu-1}(r)\phi_{\rm c}^{\nu}(r')
\right]+\cdots\,.
\end{eqnarray}
The homogeneous solution $\phi_{\rm c}^{\nu}$
is expanded to $O(v^6)$ as
\begin{eqnarray}
\phi_{\rm c}^{\nu}(z) &=&
(2z)^{\nu}
\Biggl(1 -{z^2\over 2(2\ell+3)}-{\ell \epsilon \over 2 z}
+{z^4\over 8(2\ell+3)(2\ell+5)}+{(\ell^2-5\ell-10)
\epsilon z \over 4(2\ell+3)(\ell+1)}
+{\ell(\ell-1)^2 \epsilon ^2\over 4(2\ell-1) z^2}
\nonumber \\ && \quad \qquad \qquad
-{z^6\over 48(2\ell+3)(2\ell+5)(2\ell+7)}
-{(3\ell^3-27\ell^2-142\ell-136)\epsilon z^3
\over 48(\ell+1)(\ell+2)(2\ell+3)(2\ell+5)}
\nonumber \\ && \quad \qquad \qquad
-{(\ell^3-18\ell^2+17\ell-4)\epsilon^2 \over 8(2\ell-1)^2}
-{\ell (\ell-1)(\ell-2)^2 \epsilon^3 \over 24(2\ell-1)z^3}
+O(v^7)\Biggr) \,;
\label{eq:mode-func}
\\
\nu &=&
\ell-{15\ell^2+15\ell-11\over 2(2\ell-1)(2\ell+1)(2\ell+3)}\epsilon^2 \,,
\end{eqnarray}
where $z=\omega r$ and $\epsilon = 2 M \omega$,
and $z \sim v$ and $\epsilon \sim v^3$ in the post Newtonian expansion.
The solution $\phi_{\rm c}^{-\nu-1}$
can be obtained by changing $\ell \rightarrow -\ell-1$ for $\ell \not=0,1$.
Incidentally, Eq.~(\ref{eq:mode-func}) shows
the breakdown of the post-Newtonian expansion of
$\phi_{\rm c}^{-\nu-1}$ for $\ell=0$ and $1$ at the second and third PN
orders, respectively.
The Wronskian of $\phi^{\nu}_{\rm c}$
and $\phi^{-\nu-1}_{\rm c}$ becomes
\begin{eqnarray}
\omega\,W_{\ell m\omega}(\phi_{\rm c}^{\nu},\phi_{\rm c}^{-\nu-1})
=-{2\ell+1\over 2}
-{32\ell^6+96\ell^5-176\ell^4-512\ell^3+78\ell^2+350\ell-131
\over 8 (2\ell-1)^2 (2\ell+1) (2\ell+3)^2 }\epsilon^2 +O(v^7)
\,.
\end{eqnarray}

It should be noted that $\phi_c^\nu$ contains a fractional power of $z$;
$z^\nu$. Such a factor could be an obstacle against the power
expansion. However, as seen from Eq.~(\ref{eq:gsym}),
since $z^\nu$ is associated with ${z'}{}^{-\nu-1}$
(or $z^{-\nu-1}$ with $z'{}^\nu$), the fractional power appears
always in the form $(z/z')^\nu=(r/r')^\nu$ (or $(z'/z)^\nu=(r'/r)^\nu$).
Therefore it may be expanded in powers of $(r-r')$, hence of $r$.

To determine up to which order of $(r-r')$ we have to expand
for the 3PN calculation,
we note the fact that the $O(\epsilon^2)\left(=O({v^6})\right)$
 correction in $\nu$ will give the factor
\begin{eqnarray}
-\int d\omega \sum_m
{15\ell^2+15\ell-11 \over 2(2\ell-1)(2\ell+1)(2\ell+3)}\epsilon^2
\nonumber
\end{eqnarray}
in front of the powers of $(r-r')$.
Since $\omega = m v^t/v^\phi$ under the
stationary phase approximation (which is exact for a circular orbit),
the above term behaves as $\approx \ell$
after integrating over $\omega$ and taking the $m$-sum.
The divergence after summing over $\ell$ will therefore be
$\propto (r-r')^{-2}$. Thus the expansion to $O\left({(r-r')^3}\right)$
will be sufficient,
\begin{eqnarray}
&& \left({r\over r'}\right)
^{-{15\ell^2+15\ell-11 \over 2(2\ell-1)(2\ell+1)(2\ell+3)}\epsilon^2}
\nonumber \\ && \qquad
=1-{15\ell^2+15\ell-11 \over 2(2\ell-1)(2\ell+1)(2\ell+3)}\epsilon^2
\left\{{r-r'\over r'}-{1\over 2}\left({r-r'\over r'}\right)^2
+{1\over 3}\left({r-r'\over r'}\right)^3+O\left({(r-r')^4}\right)\right\}.
\end{eqnarray}
The coefficients of the power expansion for the radial Green function
$g_{\ell m\omega}^{\rm(sym)}$ are summarized in Appendix \ref{app:coef}.
We only note here that the leading term of the post-Newtonian expansion
in $g_{\ell m\omega}^{\rm(sym)}$ behaves as $1/r$ or $1/r'$.

\subsection{Radiative part for $\ell \geq 2$}\label{app:L2H}

To the 3PN order, the post-Newtonian orders of
$\{\tilde\beta_\nu,\,\tilde\gamma_\nu\}$ shown in Eq.~(\ref{eq:betagamma})
imply that only the term proportional to
$\phi_{\rm c}^{\nu}(r)\phi_{\rm c}^{\nu}(r')$
contributes to the radiative Green function $g_{\ell m\omega}^{\rm (rad)}$,
\begin{eqnarray}
g_{\ell m\omega}^{\rm (rad)}(r,r') =
{\Im[\tilde{\gamma}_{\nu}]\,\phi_{\rm c}^{\nu}(r)\phi_{\rm c}^{\nu}(r')
\over W_{\ell m\omega}(\phi_{\rm c}^{\nu},\phi_{\rm c}^{-\nu-1})}
+\cdots\,.
\end{eqnarray}
The leading order term is then
\begin{eqnarray}
g_{\ell m\omega}^{\rm (rad)}(r,r')
= i\,{(-1)^\ell\,2^{2\ell+1}\,\omega^{2\ell+1}\,r^\ell \,r'{}^\ell
 \over 2\ell+1}\left(1+O(v^2)\right) \,,
\end{eqnarray}
which is $O(v^{2\ell+1})$ relative to the leading term in
$g_{\ell m\omega}^{\rm (sym)}$.
Hence only the mode $\ell=2$ is necessary to the 3PN order.

It may be noted that under the expansion to a finite
post-Newtonian order, only a finite number of $\ell$ contributes
to the radiative Green function.
Note also that because $g_{\ell m\omega}^{\rm (rad)}$ is pure imaginary
the radiative Green function does not contribute to the $r$-component
of the reaction force in the present case of circular orbit
around a black hole, whereas the $t$- and $\phi$-components of
the force are due to this part. This is in accordance with the
fact that the radiative Green function describes the energy and
angular momentum loss.

\subsection{$\ell = 0$ and $1$ modes}\label{app:L01}

For the modes $\ell=0,1$, since the homogeneous function
$\phi_{\rm c}^{-\nu-1}$ cannot be used,
we explicitly perform the post-Newtonian expansion
of $\phi_{\rm in}^{\nu(\ell)}$ and $\phi_{\rm up}^{\nu(\ell)}$
as done in \cite{Sasaki,PoiSas,MSSTT}.
 In the limit $\epsilon \rightarrow 0$, we have
\begin{eqnarray}
{}^{(0)}\phi_{\rm in}^{\nu(0)}(z) &=& j_0(z)
= 1-{z^2 \over 6}+{z^4 \over 120}-{z^6 \over 5040} +O(v^7)
\,, \\
{}^{(0)}\phi_{\rm in}^{\nu(1)}(z) &=& j_1(z)
= {z \over 3}
\left(1-{z^2 \over 10}+{z^4 \over 280}-{z^6 \over 15120}+O(v^7)\right)
\,, \\
{}^{(0)}\phi_{\rm up}^{\nu(0)}(z) &=& h^{(1)}_0(z)
= {-i \over z}
\left(1+i\,z-{z^2 \over 2}-{i\,z^3 \over 6}
+{z^4 \over 24}+{i\,z^5 \over 120}-{z^6 \over 720}+O(v^7)\right)
\,, \\
{}^{(0)}\phi_{\rm up}^{\nu(1)}(z) &=& h^{(1)}_1(z)
= {-i \over z^2}
\left(1+{z^2 \over 2}+{i\,z^3 \over 3}
-{z^4 \over 8}-{i\,z^5 \over 30}+{z^6 \over 144}+O(v^7)\right)
\,,
\end{eqnarray}
where $j_n$ and $h^{(1)}_n$ are the spherical Bessel function
and the spherical Hankel function of the 1st kind, respectively.
Starting from the above, the expansion to the 3PN order, that is,
to $O(\epsilon^2)$, is easily done to give
\begin{eqnarray}
\phi_{\rm in}^{\nu(0)}(z) &=&
1-{{z}^{2}\over 6}+{\frac {{z}^{4}}{120}}-{\frac {{z}^{6}}{5040}}+
\epsilon\,\left (-{5\,z \over 6}+{\frac {17\,{z}^{3}}{180}}\right )
-{\frac {11}{6}}\,{\epsilon}^{2}\ln (z) +O(v^7)
\,, \\
\phi_{\rm in}^{\nu(1)}(z) &=&
{z \over 3}\left (1-{{z}^{2} \over 10}
+{\frac {{z}^{4}}{280}}-{\frac {{z}^{6}}{15120}}
-{1 \over 2}\,{\frac {\epsilon}{z}}-{\frac {7\,z\epsilon}{20}}
+{\frac {151\,{z}^{3}\epsilon}{5040}}-{\frac {19}{30}}\,{\epsilon}^{2}\ln (z)
+O(v^7) \right )
\,, \\
\phi_{\rm up}^{\nu(0)}(z) &=&
{-i \over z}\Biggr(
1+iz-{{z}^{2} \over 2}-{i{z}^{3} \over 6}
+{{z}^{4} \over 24}+{\frac {i{z}^{5}}{120}}
-{\frac {{z}^{6}}{720}}
+{1 \over 2}\,{\frac {\epsilon}{z}}-{\frac {34\,i\epsilon}{15}}
-{\frac {7\,\epsilon\,z}{24}}+{3\,i\epsilon\,{z}^{2} \over 10}
-{{z}^{3}\epsilon \over 12}
\nonumber \\ && \qquad
+{\frac {{\epsilon}^{2}}{3\,{z}^{2}}}-{\frac {17\,i{\epsilon}^{2}}{15\,z}}
+{\frac {{\epsilon}^{3}}{4\,{z}^{3}}}+{\frac {11}{6}}\,{\epsilon}^{2}\ln (z)
-2\,z\epsilon\,\ln (z)+{\epsilon\,{z}^{3} \over 3}\ln (z)
+O(v^7) \Biggl)
\,, \\
\phi_{\rm up}^{\nu(1)}(z) &=&
{-i \over z^2}
\Biggl(1+{z^2 \over 2}+{i\,z^3 \over 3}-{z^4 \over 8}
-{i\,z^5 \over 30}+{z^6 \over 144}+{\frac {\epsilon}{z}}
-{i\epsilon\,{z}^{2} \over 6}+\epsilon\,z
+{\frac {9\,{\epsilon}^{2}}{10\,{z}^{2}}}
+{\frac {4\,{\epsilon}^{3}}{5\,{z}^{3}}}
\nonumber \\ && \qquad
+{\frac {19}{30}}\,{\epsilon}^{2}\ln (z)-{2\,{z}^{3}\epsilon \over 3}\ln (z)
+O(v^7) \Biggr)
\,.
\end{eqnarray}
The Wronskians $W_{\ell m\omega}
=W_{\ell m\omega}(\phi_{\rm in}^{\nu(\ell)},\phi_{\rm up}^{\nu(\ell)})$
 for $\ell=0$ and $1$ become
\begin{eqnarray}
\omega\,W_{0m\omega}
 &=& {i \over 180}(-180+408\,i\,\epsilon+935\,\epsilon^2)
 +O(v^7) \,, \\
\omega\,W_{1m\omega}
 &=& {i \over 225}(-225+311\,\epsilon^2) +O(v^7) \,.
\end{eqnarray}

{}From the above formulas, we find the full radial Green function
 for $\ell=0$ and $1$ as
\begin{eqnarray}
g_{0 m\omega}(r,r') &=&
{\frac {6\,{M}^{3}+3\,{r_<^3}+3\,{r_<^2}M+4\,{r_<}{M}^{2}}{3\,{r_<^4}}}
+i\omega
\nonumber \\ &&
-{\frac {1}{180\,{r_<^3}}}
\,\Bigl(300\,{M}^{2}r_>\,r_<+90\,{r_<^4}
+30\,{r_>^2}r_<\,M-3740\,{M}^{2}{r_<^2}+921\,{r_<^3}M
+1320\,{M}^{2}{r_<^2}\ln(\omega r_>)
\nonumber \\ && \qquad \qquad
+300\,Mr_>\,{r_<^2}+30\,{r_>^2}{r_<^2}+720\,{r_<^3}M\ln(\omega r_<)
+40\,{r_>^2}{M}^{2}-1320\,{M}^{2}\ln(\omega r_<)\,{r_<^2}
\Bigr){\omega}^{2}
\nonumber \\ &&
-{i \over 6}\,\left ({r_<^2}+{r_>^2}+10\,Mr_<+10\,Mr_>\right ){\omega}^{3}
\nonumber \\ &&
+{\frac {1}{360\,{r_<}^2}}
\,\Bigl(
30\,{r_>^2}{r_<^3}+307\,{r_>^2}{r_<^2}M
+15\,{r_<^5}+240\,{r_>^2}{r_<^2}M\ln(\omega r_<)
\nonumber \\ && \qquad \qquad
+300\,Mr_>\,{r_<^3}+240\,{r_<^4}M\ln(\omega r_<)
+3\,{r_>^4}M+68\,M{r_>^3}r_<+212\,{r_<^4}M+3\,{r_>^4}r_<
\Bigr){\omega}^{4}
\nonumber \\ &&
+{\frac {1}{360}}\,i\left
(3\,{r_<^4}+3\,{r_>^4}+10\,{r_>^2}{r_<^2}\right ){\omega}^{5}
-{\frac {1}{5040}}\,
{\frac {35\,{r_>^2}{r_<^4}+7\,{r_<^6}+21\,{r_>^4}{r_<^2}+{r_>^6}}{r_<}}
{\omega}^{6}
\,, \\
g_{1 m\omega}(r,r') &=&
{\frac {-10\,Mr_>\,{r_<^2}-18\,{M}^{2}r_>\,r_<+18\,{M}^{3}r_<-5\,r_>\,{r_<^3}+
5\,{r_<^3}M-32\,r_>\,{M}^{3}+10\,{M}^{2}{r_<^2}}{15\,{r_<^5}}}
\nonumber \\ &&
-{\frac {1}{1350\,{r_<^4}}}
\,\Bigl(
630\,{r_>^2}r{M}^{2}-225\,r_>\,{r_<^4}+1140\,r_>\,{M}^{2}{r_<^2}\ln(\omega r_>)
+225\,{r_<^4}M+900\,{M}^{2}{r_<^3}
\nonumber \\ && \qquad \qquad
+162\,{r_>^3}{M}^{2}-2488\,{M}^{2}r_>\,{r_<^2}
+45\,{r_>^3}{r_<^2}-1140\,r_>\,{M}^{2}\ln(\omega r_<)\,{r_<^2}+90\,M{r_>^3}r
\nonumber \\ && \qquad \qquad
+315\,{r_>^2}{r_<^2}M-900\,Mr_>\,{r_<^3}
\Bigr){\omega}^{2}
-{i \over 9}\,\left (-r_>\,r_<+Mr_<+Mr_> \right ){\omega}^{3}
\nonumber \\ &&
+{\frac {1}{7560\,{r_<^3}}}
\,\Bigl(315\,M{r_<^5}-882\,M{r_>^2}{r_<^3}-126\,{r_>^3}{r_<^3}
+9\,{r_>^5}r_<+18\,{r_>^5}M-3360\,r_>\,M\ln(\omega r_<)\,{r_<^4}
\nonumber \\ && \qquad \qquad
+151\,M{r_>^4}r_<-315\,r_>\,{r_<^5}-504\,{r_>^3}{r_<^2}M
\Bigr){\omega}^{4}
-{\frac {1}{90}}\,i\,r_>\,r_<\,\left ({r_<^2}+{r_>^2}\right ){\omega}^{5}
\nonumber \\ &&
+{\frac {1}{45360}}\,{\frac {r_>\,\left (105\,{r_<^6}
+189\,{r_>^2}{r_<^4}+27\,{r_>^4}{r_<^2}-{r_>^6}\right )}{{r_<^2}}}
{\omega}^{6}
\,,
\end{eqnarray}
where $r_{>}={\rm max}\{r,r'\}$ and $r_{<}={\rm min}\{r,r'\}$.

\section{Summary of Power Expansion Coefficients} \label{app:coef}

In this appendix, we summarize the power expansion of
the time-symmetric part of the radial
Green function $g^{\rm (sym)}_{\ell m\omega}$. It is expanded as
\begin{eqnarray}
g^{\rm (sym)}_{\ell m\omega}(r,r';r_0)
 &=& \sum_n
\left(\theta(r'-r){r^{n+\ell}\over r'^{n+\ell+1}}
 g^{\rm in(n)}_{\ell m\omega}(r',r_0)
+\theta(r-r'){r'^{n+\ell}\over r^{n+\ell+1}}
 g^{\rm out(n)}_{\ell m\omega}(r',r_0)
\right) \,,
\end{eqnarray}
where the coefficients necessary to the 3PN order calculation become
\begin{eqnarray}
g^{{\rm in}(-3)}_{\ell m\omega}(r',r_0) &=&
{(\ell ^3-5\ell ^2+8\ell -4)\ell \over 3(2\ell -1)(2\ell +1)}{M^3\over r'^3}
\,, \\
g^{{\rm in}(-2)}_{\ell m\omega}(r',r_0) &=&
-{(\ell -1)^2 \ell \over (2\ell -1)(2\ell+1)}{M^2\over r'^2}
-{(\ell -1)^2 \ell \over 2 (2\ell -1)^2 (2\ell +1)}(\omega M)^2
-{\ell (\ell ^3-\ell ^2-\ell +1)\over (2\ell -1)(2\ell+1)}{M^3\over r'^3}
\,, \\
g^{{\rm in}(-1)}_{\ell m\omega}(r',r_0) &=&
{\ell \over 2\ell +1}{M\over r'}
+{\ell \over 2(2\ell -1)(2\ell+1)}(\omega r')(\omega M)
+{\ell (\ell +1) \over 2\ell +1}{M^2\over r'^2}
+{\ell \over 8(2\ell -3)(2\ell -1)(2\ell+1)}(\omega r')^3(\omega M)
\nonumber \\ &&
+{\ell ^2+7\ell -4\over 2 (2\ell -1)(2\ell+1)}(\omega M)^2
+{\ell (\ell ^3+5\ell ^2+8\ell +4)\over (2\ell +1)(2\ell+3)}{M^3\over r'^3}
\,, \\
g^{{\rm in}(+0)}_{\ell m\omega}(r',r_0) &=&
-{1\over 2\ell +1}
-{1\over 2(2\ell -1)(2\ell+1)}(\omega r')^2
-{\ell +1\over 2\ell +1}{M\over r'}
-{1\over 8(2\ell -3)(2\ell -1)(2\ell+1)}(\omega r')^4
\nonumber \\ &&
-{\ell ^2+7\ell -4\over 2\ell (2\ell -1)(2\ell+1)}(\omega r')(\omega M)
-{(\ell+1)(\ell+2)^2 \over (2\ell +1)(2\ell +3)}{M^2\over r'^2}
-{1\over 48(2\ell -5)(2\ell -3)(2\ell -1)(2\ell+1)}(\omega r')^6
\nonumber \\ &&
-{3\ell ^3+36\ell ^2-79\ell +24\over
 24 (\ell -1)\ell (2\ell-3)(2\ell -1)(2\ell+1)}(\omega r')^3(\omega M)
\nonumber \\ &&
-{1488\ell^6+7104\ell^5+10608\ell^4+1940\ell^3-6825\ell^2-2321\ell+1740
\over 6 (2\ell-1)^2 (2\ell+1)^3 (2\ell+3)^2 }
(\omega M)^2
\nonumber \\ &&
+{2(15\ell ^2+15\ell -11)\over (2\ell -1) (2\ell +1)^2(2\ell +3)}
(\omega M)^2 \ln(r_0/r')
-{\ell ^4+9\ell ^3+29\ell ^2+39\ell +18\over 3(2\ell +1)(2\ell +3)}
{M^3\over r'^3}
\,, \\
g^{{\rm in}(+1)}_{\ell m\omega}(r',r_0) &=&
-{\ell ^2-5\ell -10\over 2(\ell +1)(2\ell +1) (2\ell +3)}(\omega r')(\omega M)
-{\ell ^2-5\ell -10 \over 4(\ell +1)(2\ell -1) (2\ell +1)(2\ell +3)}
(\omega r')^3(\omega M)
\nonumber \\ &&
+{6(15\ell ^2+15\ell -11) \over (2\ell -1) (2\ell +1)^2(2\ell +3)}
{r'\over r_0}(\omega M)^2
-{\ell ^2-5\ell -10 \over 2(2\ell +1)(2\ell +3)}(\omega M)^2
\,, \\
g^{{\rm in}(+2)}_{\ell m\omega}(r',r_0) &=&
{1\over 2(2\ell +1)(2\ell +3)}(\omega r')^2
+{1\over 4(2\ell -1)(2\ell +1)(2\ell +3)}(\omega r')^4
+{\ell +1\over 2(2\ell +1)(2\ell +3)}(\omega r')(\omega M)
\nonumber \\ &&
+{1\over 16(2\ell -3)(2\ell -1)(2\ell +1)(2\ell +3)}(\omega r')^6
+{\ell ^2+7\ell -4\over 4\ell (2\ell -1)(2\ell +1)(2\ell +3)}
(\omega r')^3(\omega M)
\nonumber \\ &&
+{3(15\ell^2+15\ell-11) \over (2\ell -1)(2\ell +1)^2(2\ell +3)}
{r'^2\over r_0^2}(\omega M)^2
-{(\ell+1) (\ell+2)^2 \over 2 (2\ell +1)(2\ell +3)^2}(\omega M)^2
\,, \\
g^{{\rm in}(+3)}_{\ell m\omega}(r',r_0) &=&
{3\ell ^3-27\ell ^2-142\ell -136\over
24(2\ell ^3+11\ell ^2+19\ell +10)(2\ell +3)(2\ell +1)}
(\omega r')^3(\omega M)
\nonumber \\ &&
+{2 (15\ell^2+15\ell -11) \over 3 (2\ell -1)(2\ell +1)^2(2\ell +3)}
\left({r' \over r_0}\right)^3 (\omega M)^2
\,, \\
g^{{\rm in}(+4)}_{\ell m\omega}(r',r_0) &=&
-{1\over 8(2\ell +1)(2\ell +3)(2\ell +5)}(\omega r')^4
-{1\over 16(2\ell -1)(2\ell +1)(2\ell +3)(2\ell +5) }(\omega r')^6
\nonumber \\ &&
-{\ell +1\over 8(2\ell +1)(2\ell +3)(2\ell +5)}(\omega r')^3(\omega M)
\,, \\
g^{{\rm in}(+5)}_{\ell m\omega}(r',r_0) &=&
0
\,, \\
g^{{\rm in}(+6)}_{\ell m\omega}(r',r_0) &=&
{1\over 48(2\ell +7)(2\ell +1)(2\ell +3)(2\ell +5)}(\omega r')^6
\,, \\
g^{{\rm out}(+3)}_{\ell m\omega}(r',r_0) &=&
-{\ell ^4+9\ell ^3+29\ell ^2+39\ell +18
\over 3(2\ell +1)(2\ell +3)}{M^3\over r'^3}
\,, \\
g^{{\rm out}(+2)}_{\ell m\omega}(r',r_0) &=&
-{(\ell+1)(\ell+2)^2 \over (2\ell +1)(2\ell +3)}{M^2\over r'^2}
+{(\ell+1)(\ell+2)^2 \over 2(2\ell +1)(2\ell +3)^2}(\omega M)^2
+{\ell(\ell+1)(\ell+2)^2 \over (2\ell +1)(2\ell +3)}{M^3\over r'^3}
\,, \\
g^{{\rm out}(+1)}_{\ell m\omega}(r',r_0) &=&
-{\ell +1\over 2\ell +1}{M\over r'}
+{\ell +1\over 2 (2\ell +1)(2\ell +3)}(\omega r')(\omega M)
+{\ell (\ell +1) \over 2\ell +1}{M^2\over r'^2}
\nonumber \\ &&
-{\ell +1 \over 8 (2\ell +1)(2\ell +3)(2\ell +5)}(\omega r')^3(\omega M)
-{\ell ^2-5\ell -10\over 2 (2\ell +1)(2\ell +3)}(\omega M)^2
-{(\ell-1)^2 \ell (\ell+1) \over (2\ell -1)(2\ell +1)}{M^3\over r'^3}
\,, \\
g^{{\rm out}(+0)}_{\ell m\omega}(r',r_0) &=&
-{1\over 2\ell +1}
+{1\over 2(2\ell +1)(2\ell +3)}(\omega r')^2
+{\ell \over 2\ell +1}{M\over r'}
-{1\over 8(2\ell +1)(2\ell +3)(2\ell +5)}(\omega r')^4
\nonumber \\ &&
-{\ell ^2-5\ell -10 \over 2 (\ell +1)(2\ell +1)(2\ell +3)}(\omega r')(\omega M)
-{\ell(\ell -1)^2 \over (2\ell -1)(2\ell +1)}{M^2\over r'^2}
\nonumber \\ &&
+{1\over 48(2\ell +1)(2\ell +3)(2\ell +5)(2\ell +7)}(\omega r')^6
+{3\ell ^3-27\ell ^2-142\ell -136\over
24(\ell +1)(\ell +2)(2\ell +1)(2\ell +3)(2\ell +5)}
(\omega r')^3(\omega M)
\nonumber \\ &&
-{1488\ell^6+1824\ell^5-2592\ell^4-788\ell^3+2283\ell^2-1309\ell+288 \over
6 (2\ell -1)^2(2\ell +1)^3(2\ell +3)^2}(\omega M)^2
\nonumber \\ &&
+{2(15\ell ^2+15\ell -11)\over (2\ell -1)(2\ell +1)^2 (2\ell +3)}
(\omega M)^2\ln(r'/r_0)
+{(\ell ^3-5\ell ^2+8\ell -4)\ell \over 3(2\ell +1)(2\ell -1)}{M^3\over r'^3}
\,, \\
g^{{\rm out}(-1)}_{\ell m\omega}(r',r_0) &=&
-{\ell ^2+7\ell -4\over 2\ell (2\ell-1)(2\ell+1) }(\omega r')(\omega M)
+{\ell ^2+7\ell -4\over 4 (2\ell-1)(2\ell+1)(2\ell+3)}(\omega r')^3(\omega M)
\nonumber \\ &&
-{6(15\ell^2+15\ell-11) \over (2\ell-1)(2\ell+1)^2(2\ell+3)}
{r'\over r_0}(\omega M)^2
+{\ell ^2+7\ell -4\over 2 (2\ell-1)(2\ell+1) }(\omega M)^2
\,, \\
g^{{\rm out}(-2)}_{\ell m\omega}(r',r_0) &=&
-{1\over 2 (2\ell-1)(2\ell+1) }(\omega r')^2
+{1\over 4 (2\ell-1)(2\ell+1)(2\ell+3)}(\omega r')^4
+{\ell \over 2(2\ell-1)(2\ell+1)}(\omega r')(\omega M)
\nonumber \\ &&
-{1\over 16(2\ell-1)(2\ell+1)(2\ell+3)(2\ell+5)}(\omega r')^6
-{\ell ^2-5\ell -10 \over 4(\ell +1)(2\ell-1)(2\ell+1)(2\ell+3)}
(\omega r')^3(\omega M)
\nonumber \\ &&
+{3(15\ell^2+15\ell-11) \over (2\ell-1)(2\ell+1)^2(2\ell+3)}
\left({r'\over r_0}\right)^2(\omega M)^2
\nonumber \\ &&
+{(\ell-1)^2\ell \over 2 (2\ell-1)^2(2\ell+1)}(\omega M)^2
\,, \\
g^{{\rm out}(-3)}_{\ell m\omega}(r',r_0) &=&
-{3\ell ^3+36\ell ^2-79\ell +24\over
24 (\ell -1)\ell (2\ell -3)(2\ell -1)(2\ell +1)}
(\omega r')^3(\omega M)
\nonumber \\ &&
-{2 (15\ell^2+15\ell-11) \over 3 (2\ell-1)(2\ell+1)^2(2\ell+3)}
\left({r' \over r_0}\right)^3(\omega M)^2
\,, \\
g^{{\rm out}(-4)}_{\ell m\omega}(r',r_0) &=&
-{1\over 8(2\ell -3)(2\ell -1)(2\ell +1)}(\omega r')^4
+{1\over 16(2\ell -3)(2\ell -1)(2\ell +1)(2\ell +3)}(\omega r')^6
\nonumber \\ &&
+{\ell \over 8(2\ell -3)(2\ell -1)(2\ell +1)}(\omega r')^3(\omega M)
\,, \\
g^{{\rm out}(-5)}_{\ell m\omega}(r',r_0) &=&
0
\,, \\
g^{{\rm out}(-6)}_{\ell m\omega}(r',r_0) &=&
-{1\over 48(2\ell -5)(2\ell +1)(2\ell -1)(2\ell -3)}(\omega r')^6
\,.
\end{eqnarray}

\section{Generating function of $m$ sum}\label{app:m-sum}

In this appendix,
we describe a method to sum over the $m$-modes of spherical harmonics for
arbitrary $\ell$ that appears in Eq.~(\ref{eq:full-power-r0}).
Specifically, the $m$-sum we need to evaluate takes the form,
\begin{eqnarray}
\sum_{m=-\ell}^\ell m^N|Y_{\ell m}(\pi/2,0)|^2\,,
\label{eq:msumform}
\end{eqnarray}
where $N$ is a non-negative integer. To perform the above summation, we
introduce the generating function,
\begin{eqnarray}
\Gamma_{\ell}(z)
 =\sum_{m=-\ell}^{\ell} e^{mz} |Y_{\ell m}(\pi/2,0)|^2 \,.
\end{eqnarray}
Then the sum (\ref{eq:msumform}) may be evaluated as
$\lim\limits_{z\to0}\partial_z^N\Gamma_{\ell}(z)$.

First note that $Y_{\ell m}(\pi/2,0)$ is non-zero only
when $\ell-m$ is an even integer. So putting $2n=\ell-m$, we have
\begin{eqnarray}
\Gamma_{\ell}(z)
&=& \sum_{n=0}^{\ell} e^{(l-2n)z}
{2\ell+1\over 4\pi}{(2n)!\over (2\ell-2n)!}
\left({(2\ell-2n-1)!!\over (2n)!!}\right)^2
\nonumber\\
&=& {2\ell+1 \over 4\pi}{e^{\ell z}\over 4^{\ell}} \sum_{n=0}^{\ell}
{\Gamma(2n+1)\Gamma(2\ell-2n+1)\over \Gamma(n+1)\Gamma(\ell-n+1)^2}
{e^{-2nz}\over n!}
\nonumber\\
&=& {2\ell+1\over 4\pi^2}
{\Gamma(\ell+1/2)\Gamma(-\ell+1/2)\over \Gamma(\ell+1)\Gamma(-\ell)}
 e^{\ell z}
\sum_{n=0}^{\infty}
{\Gamma(n+1/2)\Gamma(n-\ell)\over \Gamma(n-\ell+1/2)}{e^{-2nz}\over n!}
\nonumber\\
&=& {2\ell+1\over 4\pi^2}
{\Gamma(\ell+1/2)\Gamma(1/2)\over \Gamma(\ell+1)} e^{\ell z}
{}_2F_1 \left({1\over 2},\,-\ell;\,-\ell+{1\over 2};\,e^{-2z}\right) \,,
\end{eqnarray}
where ${}_2F_1$ is the hypergeometric function,
and we have temporarily
performed the analytic continuation of $\ell$ to a complex number
from the second line to the third line.

Using the relation between the hypergeometric functions as
\begin{eqnarray}
{}_2F_1 \left({1\over 2},\,-\ell;\,-\ell+{1\over 2};\,x\right) &=&
(-1)^{\ell}{\Gamma(\ell+1)\Gamma(-\ell+1/2)\over \Gamma(1/2)}
{}_2F_1 \left({1\over 2},\,-\ell;\,1;\,1-x\right) \,,
\end{eqnarray}
for a non-negative integer $\ell$, we have
\begin{eqnarray}
\Gamma_{\ell}(z) &=& {2\ell+1\over 4\pi}e^{\ell z}
{}_2F_1 \left({1\over 2},\,-\ell;\,1;\,1-e^{-2z}\right) \,.
\end{eqnarray}
This can be easily expanded to arbitrary order of $z$.
For example, to $O(z^6)$ we have
\begin{eqnarray}
\Gamma_{\ell}(z) &=& {2\ell+1\over 4\pi}
\Biggl\{1+\left({\ell \,(\ell+1) \over 2}\right){1\over 2}z^2
+\left({\ell \,(\ell+1)(3 \ell^2+3\ell-2) \over 8}\right)
{1\over 4!}z^4
\nonumber \\ && \qquad \qquad
+\left({\ell \,(\ell+1)(5\ell^4+10\ell^3-5\ell^2-10\ell+8) \over 16}
\right){1\over 6!}z^6
+O(z^8)\Biggr\} \,.
\end{eqnarray}

\end{appendix}


\end{document}